\documentclass[author,usecmfonts,genTeX,reqno]{nrc2}
\usepackage{graphicx}
\usepackage{amsmath,amssymb,array,rotating}
\usepackage{bm}
\usepackage{multirow}
\usepackage{cite}
\journalcode{cjp}
\usepackage{verbatim}
\usepackage[title]{appendix}

\begin{document}

\title{Irreversible Thermodynamics of Transport across Interfaces}
\author{ Matthew R. Sears }
\address[TAMU]{ Department of Physics, Texas A\&M University, College Station, TX 77843-4242}
\author{ Wayne M. Saslow }
\address[TAMU]
\correspond{wsaslow@tamu.edu}
\shortauthor{Sears, Saslow}

\begin{abstract}
With spintronics applications in mind, we use irreversible thermodynamics to derive the rates of entropy production and heating near an interface when heat current, electric current, and spin current cross it.  Associated with these currents are apparent discontinuities in temperature $(\Delta T)$, electrochemical potential $(\Delta \tilde{\mu})$, and spin-dependent ``magnetoelectrochemical potential'' $(\Delta {\bar{\mu}_{\uparrow,\downarrow}})$.   This work applies to magnetic semiconductors and insulators as well as metals, due to the inclusion of the chemical potential $\mu$, which usually is neglected in works on interfacial thermodynamic transport.  We also discuss the (non-obvious) distinction between entropy production and heat production.  Heat current and electric current are conserved, but spin current is not, so it necessitates a somewhat different treatment.  At low temperatures or for large differences in material properties, the surface heating rate dominates the bulk heating rate near the surface.  We also consider the case, noted by Rashba, where bulk spin currents occur in equilibrium.  Although a surface spin current (in A/m$^{2}$) should yield about the same rate of heating as an equal surface electric current, production of such a spin current requires a relatively large ``magnetization potential'' difference across the interface. 

\PACS{05.70.Ln,05.70.Np,67.40.Pm,73.40.Cg}
\end{abstract}

\begin{resume}
Avec applications dans l'esprit de spintronics, nous employons la thermodynamique irr\'{e}versible \`{a} obtenir les taux de production d'entropie et de chauffage \`{a} proximit\'{e} d'une interface lorsque la chaleur actuelle, le courant \'{e}lectrique, et courant de spin la traverser.
Associ\'{e}s \`{a} ces courants sont discontinuit\'{e}s apparentes de la temp\'{e}rature $ (\Delta T) $, potentiel \'{e}lectrochimique $ (\Delta \tilde {\mu}) $, et d\'{e}pendant du spin potentiel ``magneto\'{e}lectrochimique'' $ (\Delta{\bar{\mu}_{\uparrow, \downarrow}})$.
Ce travail s'applique \`{a} semi-conducteurs magn\'{e}tiques et isolants ainsi que des m\'{e}taux, due \`{a} l'inclusion de la potentiel chimique $\mu$, ce qui est g\'{e}n\'{e}ralement n\'{e}glig\'{e}e dans les travaux sur les transports thermodynamique interfaciale. Nous discutons \'{e}galement de la distinction (non \'{e}vidente) entre la production d'entropie et la production de chaleur. Chaleur actuelle et le courant \'{e}lectrique sont conserv\'{e}s, mais n'est pas courant de spin, il n\'{e}cessite un traitement quelque peu diff\'{e}rent. A basse temp\'{e}rature, ou pour de grandes diff\'{e}rences dans les propri\'{e}t\'{e}s du mat\'{e}riau, la vitesse de chauffage de surface domine la vitesse de chauffage en vrac pr\'{e}s de la surface. Nous consid\'{e}rons \'{e}galement le cas, a not\'{e} par Rashba, o\'{u} les courants de spin en vrac se produire \`{a} l'\'{e}quilibre. M\^{e}me si un courant de spin de surface (en A/m$ ^ {2} $) devrait donner environ le m\^{e}me taux de chauffage d'une surface \'{e}gale de courant \'{e}lectrique, la production d'un tel courant de spin n\'{e}cessite un potentiel relativement important ``aimantation diff\'{e}rence'' entre l'interface.
\end{resume}

\maketitle

\section{Introduction}

It is well-known that apparent voltage and temperature discontinuities, determined by extrapolation from the bulk, appear at interfaces in the presence of heat or electric current.  For small currents, these discontinuities are proportional to the heat or electric current.  For heat current, the coefficient of proportionality is known as the thermal boundary resistance, and was first studied at low temperatures by Kapitza for the solid--liquid $^4$He interface \cite{Khalat, Pollack, Challis, SwartzPohl}.  For electric current, the coefficient of proportionality is known as the surface resistance, or specific resistance \cite{BassPratt07}.  In principle, there can also be off-diagonal terms, corresponding to a discontinuity in the temperature causing an electric current \cite{JohnsonSilsbee}.  There also are 
more recently, spin-dependent conduction effects across surfaces, as studied, for example, in \cite{JohnsonSilsbee,vSonvKempWyder,ValetFert}.    

Johnson and Silsbee \cite{JohnsonSilsbee} studied the surface and bulk transport coefficients for these phenomena, but without considering details of the non-conservation of the spin current due to spin-flip processes, and did not study the rate of heating near the surface.  
The present work considers these non-conservation phenomena, which require more refined considerations than when they are not present.  (The terms ``magnetoelectrochemical potential'' and ``magnetization potential'' that appear in the theory  were first employed in \cite{JohnsonSilsbee}; they are made more precise below.) The present work also includes the chemical potential $\mu$ of the charge-carriers, neglected by \cite{JohnsonSilsbee}, which considered metals.  The present results are more general; $\mu$ is negligible (compared to electrical potential energy) for metals but not necessarily for semiconductors or insulators, where small changes in carrier density can have a large effect on $\mu$.

The present work also considers the conditions under which bulk heating dominates surface heating, and vice-versa.  It also explicitly considers the near-surface region.  Note that heating implies entropy production, but not the converse; Sect. IIB presents some considerations on this matter.

For specificity, note that if there is an extrapolated temperature discontinuity $\Delta T$ across a solid-solid interface, the energy flux (and heat flux) are given by 
\begin{equation}
|j^{\varepsilon}|=h_{K} |\Delta T|,
\end{equation}
where $h_{K} \ge0$ is the Kapitza, or thermal boundary, conductance and $R_{K}=h_{K}^{-1}$ is the Kapitza, or thermal boundary, resistance.  Normally this expression is simply written down, \cite{JohnsonSilsbee} being an exception (with a very different notation).  (As usual, heat flows from hot to cold, and we assume no voltage drop across the interface.)  Entropy is produced but no heating occurs, as thermal equilibration occurs by heat flow. 

In addition, when there is an extrapolated voltage discontinuity $\Delta V$ across a solid-solid interface, the electric current flux is given by
\begin{equation}
|j|=\bar{g}|\Delta V|,
\label{j_surf}
\end{equation}
where $\bar{g}$ is the surface electrical conductance.  Likewise, this equation normally is simply written down, although we note \cite{JohnsonSilsbee, vSonvKempWyder,ValetFert}.  To our knowledge, in none of these cases is the relationship to heating made.  (As usual, current flows from high to low voltage, and we assume no temperature drop across the interface.)  In this case there is both heating and entropy production. 

Although both heat current and electric current are associated with a rate of entropy production ${\cal S}_{s}$, only the latter is associated with the rate of heat production.  Specifically, the rate of heating (per unit area) associated with an interface is given only by 
\begin{equation}
{\cal R}_I =|j(\Delta V)|,
\label{R_def}
\end{equation}
a simple generalization of the bulk rate of heating (per unit volume) of $\vec{j}\cdot\vec{E}$.  Furthermore, when there also is a spin current $J^{\sigma}$ (in units of 1/s), there is an additional rate of heating (per unit area) associated with an interface, given below, as the third term in eq.\eqref{2R_S1}.

The fundamental principle that ensures \eqref{R_def} is the non-decreasing nature of the entropy of the overall system, just as that same principle ensures the corresponding relation \eqref{j_surf} between electric current and voltage gradient; both yield current flow from high to low voltage.  

General references on irreversible thermodynamics are \cite{deGrootMazur,Prigogine}, and appropriate sections in the thermodynamics texts by Callen and by Morse \cite{Callen,Morse}. A number of more recent approaches and applications, closer to the spirit of the present work, are also available \cite{HalpHohen,FLMSP,MartinParodiPershan,Graham,HuSaslow,Liu-mag,Forster,Reichl}.  The present work does not consider length scales so small that ordinary heat conduction (e.g., the Fourier law) is not expected to hold, due either to classical or quantum size effects \cite{MahanMaris}.  
Although very general, irreversible thermodynamics only applies to systems near  equilibrium; it does not treat situations like the free expansion of a gas.

The present work does not consider systems for which the interface is a distinct thermodynamic system.  For a system that is out of thermodynamic equilibrium a surface temperature may not be a well-defined quantity.  (See \cite{Lumpkin1,Maiti} for molecular dynamics simulations of heat flow across an interface, which show a sharp temperature jump at the atomic level.)   Moreover, thermometers that measure different properties, but are calibrated in the bulk, need not read equivalent temperatures near the surface.  This is because near surfaces the thermal distribution function is not defined solely in terms of thermodynamic properties, but also in terms of surface solutions of the transport equation \cite{SaslowSurfSols, PennStiles1}.

Section~\ref{SingleCarrier} considers heating in a single charge-carrier system (or even a system with no charge-carriers) and obtains, in addition to the well-known bulk heating rate, the surface heating rate.  Section~\ref{Spin} considers a two charge-carrier system (with spins in mind), obtaining the bulk heating rate (including equilibrium spin currents, which, if non-uniform, must have zero divergence) and the surface heating rate.  Section~\ref{Conclusion} provides a brief summary. 

A number of additional considerations are given in the Appendices.  The approach taken in the present work employs intensive thermodynamic variables associated with the energy-maximum principle (e.g., $T$, $P$, $\mu$), and considers the case where certain of the extensive thermodynamic variables are not conserved.  On the other hand, the approach taken in \cite{Callen,Morse} employs variables associated with the entropy-minimum principle (e.g., $1/T$, $-P/T$, $-\mu/T$), but considers only the case where all of the extensive thermodynamic variables are conserved.  Appendix~\ref{Entropy-BasedAppendix} shows how to use the entropy-minimum variables to obtain the irreversible thermodynamics when the extensive thermodynamic variables are either conserved or not conserved.  Appendix~\ref{SoundAppendix} shows that, in the long-wavelength limit, the entropy of a sound wave is zero.  Appendix~\ref{NotationAppendix} compares the notation of the present work with that of \cite{JohnsonSilsbee}.  Note that Ref.~\cite{JohnsonSilsbee} considers metals, whereas the present work contains results that are also applicable to magnetic semiconductors. 

The same approach can be applied to interfaces involving electrons and holes, rather than up-spin and down-spin electrons.  Just as the spin current is not conserved, because up-spins and down-spins can flip, so too the difference between the electric currents due to electrons and holes is not conserved, due to electron-hole recombination.

\section{Single Carrier Systems}
\label{SingleCarrier}
The spirit of irreversible thermodynamics is to write down the differential of the energy density $\epsilon$ in terms of the intensive quantities (like $T$ and $\mu$) multiplied by their thermodynamically conjugate extensive densities (like the entropy density $s$ and the carrier number density $n$).  Next the equations of motion (often conservation laws) for all the extensive densities are written down in terms of unknown fluxes and sources.  Then the volume rate of entropy production ${\cal S}$, which is non-zero, is written in terms of a divergence and a product of unknown fluxes with corresponding gradients of intensive parameters.  Finally, the condition that ${\cal S}\ge0$ completely determines the form of the unknown fluxes (and sources), although the values of the transport coefficients are specified only subject to certain inequalities and, for the off-diagonal terms, to the Onsager symmetry relations.  

\subsection{Rate of Entropy Production}
For a single carrier conductor, we assume that the bulk energy density $\varepsilon$, bulk number density $n$, and bulk entropy density $s$ are related by 
\begin{equation}
d\varepsilon=Tds + \tilde{\mu} dn.
\label{depsilon}
\end{equation}
The electrochemical potential $\tilde{\mu}=\mu-eV$, where $\mu$ is the chemical potential.  (In principle, a material can contain multiple heat-carrying subsystems, which may be at different temperatures \cite{SandWalt}, so that $d\varepsilon =  T_{1} ds_{1} + T_{2} ds_{2} + \dots$; see Ref.~\cite{SearsSasSSE} for a discussion of the irreversible thermodynamics of such systems.) 
The continuity relations for the conserved number and energy densities are 
\begin{eqnarray}
\frac{\partial n}{\partial t} + \partial_{i} j^n_{i}&=&0, \label{n_eq}\\
\frac{\partial \varepsilon}{\partial t} +\partial_{i} j^\varepsilon_{i} &=&0.
\label{e_eq}
\end{eqnarray}
The equation for the non-conserved entropy density is 
\begin{equation}
\frac{\partial s}{\partial t} + \partial_{i} j^s_{i} \equiv {\cal S}_{s} \ge0.
\label{R/T}
\end{equation}
Alternatively one can use the dissipation function
\begin{align}
R = T {\cal S}_{s}
\end{align}
as the primary quantity.  $R$ has the units of a rate of heating, but is only a rate of heating when energy is transformed into heat, not when it already is in the form of heat.  For a view more exclusively based on ${\cal S}_{s}$, see \cite{KjelsBed}.

Combining the above equations yields
\begin{equation}
\begin{split}
0 \leq T{\cal S}_{s} &= - \partial_{i} j^\varepsilon_{i} + \tilde{\mu} \partial_{i} j^n + T \partial_{i} j^s \\
 &= \partial_{i} \left( -  j^\varepsilon_{i} + \tilde{\mu} j^n_{i} + T j^s_{i} \right) - j^s_{i} \partial_{i} T - j^n_{i} \partial_{i}\tilde{\mu} \rm{.}
\label{R1}
\end{split}
\end{equation} 
Following the approach of irreversible thermodynamics, this has been written as a single divergence term and the sum of products of unknown fluxes with gradients of the thermodynamic intensive quantities.  (These gradients are also called thermodynamic forces.)  

Since the divergence term may be either positive or negative, it must always be zero to ensure that entropy never decreases, i.e., ${\cal S}_{s} \geq 0$.  Thus
\begin{equation}
 j^\varepsilon_{i} = \tilde{\mu} j^n_{i} + T j^s_{i}, 
 \label{jRelation}
\end{equation}
and
\begin{equation}
0 \leq T{\cal S}_{s}  =  - j^s_{i} \partial_{i} T - j^n_{i} \partial_{i} \tilde{\mu} .
\label{R2}
\end{equation} 

If there is no number current $j^{n}_{i}$, then the energy current $j^{\epsilon}_{i}$ consists only of a heat current $j^{Q}_{i}\equiv Tj^{s}_{i}$.  In this case the energy is only in the form of heat, and there is no additional heat production, although there is entropy production.  

\subsection{Bulk Fluxes and Rate of Entropy Production}
In bulk, by the non-negativity of \eqref{R2}, the linearized flux densities take the form 
\begin{eqnarray}
j^s_{i} &=& - \frac{\kappa}{T} \partial_{i} T - L_{sn} \partial_{i} \tilde{\mu} , \label{jsSingle}\\
j^n_{i} &=& - L_{ns} \partial_{i} T - \frac{\sigma}{e^2} \partial_{i} \tilde{\mu} , \label{jnSingle}
\end{eqnarray}
where $\kappa$ is the thermal conductivity, $\sigma$ is the electrical conductivity, $e$ is electric charge, and $L_{sn} = L_{ns}$ by the Onsager principle.  Irreversible thermodynamics cannot provide values for any of these material-dependent coefficients, but Kubo theory can give these coefficients in terms of equilibrium correlation functions \cite{Forster,Reichl,Mahan}.

By \eqref{R2}-\eqref{jnSingle}, the rate of entropy production in the bulk (per unit volume) is given by
\begin{equation}
{\cal S}_{s} =\frac{\kappa}{T^{2}} \left( \partial_{i} T \right)^2 + \frac{\sigma}{e^2T} \left( \partial_{i} \tilde{\mu} \right)^2  + 2 \frac{L_{sn}}{T} \left( \partial_{i} \tilde{\mu} \right) \left( \partial_{i} T \right) .
\label{R_B}
\end{equation}
By ${\cal S}_{s} \ge 0$ we have $\kappa\ge0$, $\sigma \ge 0$, and $L^{2}_{sn}\le (\sigma \kappa)/(e^2 T)$.  As noted in the introduction, pure thermal conduction already involves heat flow, so there is no production of heat in that case.  

On the other hand, the entropy production due to current flow does cause heating, at the rate per volume of 
\begin{equation}
R \equiv -j^{n}_{i}\partial_{i}\tilde{\mu} = \frac{\sigma}{e^2} \left( \partial_{i} \tilde{\mu} \right)^2+L_{sn} (\partial_i \tilde{\mu})( \partial_i T).
\label{R_B2}
\end{equation}
Only the first term is Joule heating.  The second term is like Thomson heating, in that it can have either sign.\footnote{Because Thomson heating (as discussed in Ref.~\cite{Callen}) involves the artificial maintenance of the same temperature distribution both with and without current flow, we hesitate to call this cross-term Thomson heating, although the latter involves both a temperature gradient and a voltage gradient. }  
For a large $\partial_{i} T$, it can even dominate, but the net entropy production ${\cal S}_{s}$ remains non-negative. 

\subsection{Rates of Heating and Entropy Production}
When calculating the rate of heating this elimination, by hand, of the part of $T{\cal S}_{s}$ associated with heat flow is related to a similar effect discussed in \cite{LLFluid} of damping of a sound wave.  In that case the mechanical energy $E_{mech}$ of the sound wave (which, implicitly, has zero entropy) dissipates into heat, which increases the entropy $S$ of the background system by $\dot{E}_{mech}=-T\dot{S}$.  $\dot{E}_{mech}$ is determined by a volume integration over the equivalent of $R$, evaluated for the sound wave, and is proportional to the square of the sound wave amplitude (including temperature oscillations in the sound wave).  This results in hot spots (as, for a standing wave, is perhaps familiar from a microwave oven) that separately diffuse.  However, this energy is already heat energy.  Once deposited as heat, its diffusion causes a further increase in entropy, but no additional energy goes into the system.  Note that in the long-wavelength limit the entropy per unit mass $\sigma$ is conserved, from which one can show (see Appendix \ref{SoundAppendix}) that the entropy of a sound wave is zero.  

Another example where entropy increase and heating are distinct is a gas of interacting atoms that has a multi-nanometer range for repulsion.  Let all the atoms initially be placed within an interaction volume of one another.  When they become thermally disordered, the increase in entropy can be treated as in the present work, but only the interaction energy converts into heat.   

In spin-Seebeck experiments \cite{UchidaPy,UchidaInsul,AwschMyers}, the applied thermal gradient causes heat flow between various subsystems, which in turn induces a spin current \cite{Bauer1,SearsSasSSE}.  Although it increases the entropy of the system, the heat flow is not associated with heating; however, spin and electrical currents both increase the entropy of the system and cause heating. Spin currents are addressed in Section~\ref{Spin}.

\subsection{Surface Fluxes and Rate of Entropy Production}
At low temperatures, or when the material properties change significantly on crossing the interface $I$, the changes in $T$ and $\mu$ at the interface are very large, and \eqref{R_B} integrated over the surface region (assuming that $T$ and $\tilde{\mu}$ are well-defined in this region) can be smaller than the surface entropy production rate ${\cal S}_{I}$.  In this subsection we consider ${\cal S}_{I}$.  In the next subsection we consider the conditions under which ${\cal S}_{I}$ dominates \eqref{R_B} integrated over the surface region.  This involves considerations of the characteristic mean-free-path $l$ and the distance $a$ over which the thermodynamic quantities adjust to the surface.  

The total rate (per unit area) of entropy production at the interface, ${\cal S}_{I} $, is obtained by integrating the volume rate of entropy production over the surface region.  By \eqref{R2}, taking flow only in the $x$-direction,
\begin{equation}
{\cal{S}}_{I} = \int_{I} dx {\cal S}_{s} = -\int_{I} dx \frac{\left( j^s \partial_x T + j^n \partial_x \tilde{\mu}  \right)}{T} \rm{.}
\label{TotalHeating1}
\end{equation}
In steady-state, the energy and number flux densities \eqref{jRelation} and \eqref{jnSingle} are uniform across this region.  If $T$ and $\mu$ are also nearly uniform, by \eqref{jRelation} the entropy flux density will also be nearly uniform, so  
\begin{equation}
\begin{split}
{\cal{S}}_I &\approx - \frac{j^s}{T} \int_{I} dx \partial_x T - \frac{j^n}{T} \int_{I} dx  \partial_x \tilde{\mu} \\
&\approx - \frac{j^s}{T} \left( \Delta T \right)_I -\frac{j^n}{T} \left( \Delta \tilde{\mu} \right)_{I}. 
\label{TotalHeating2}
\end{split}
\end{equation}
Here $\left( \Delta T \right)_{I}$ and $\left( \Delta \tilde{\mu} \right)_{I}$ are the differences of temperature and electrochemical potential across the interface region.  Johnson and Silsbee \cite{JohnsonSilsbee} did not include the effect of variations in the chemical potential $\mu$, which normally does not matter for metals (which motivated their work).  However, variations in $\mu$ often are of significance for semiconductors, where they cause diffusion.  On setting $\tilde\mu=\mu-eV\approx-eV$, eq.~\eqref{TotalHeating2} agrees with eq.~(12) of \cite{JohnsonSilsbee}.  See Table~\ref{JSTable} for a term-by-term comparison.

We now apply the same type of irreversible thermodynamics approach to the surface region as to the bulk.  In bulk the fluxes are proportional to the gradients of the intensive thermodynamic quantities.  At a surface, the fluxes are taken to be proportional to the differences across the interface of the intensive thermodynamic quantities.  Thus 
\begin{eqnarray}
j^s &=& -\frac{h_{K}}{T} \left( \Delta T \right)_{I} - L'_{sn} \left( \Delta \tilde{\mu} \right)_{I} , \label{js1} \\
j^n &=& -L'_{ns} \left( \Delta T \right)_{I} - \frac{\bar{g}}{e^2} \left( \Delta \tilde{\mu} \right)_{I} . \label{jn1}
\end{eqnarray}
Here, $h_{K}$ is the thermal boundary resistance, and is of the order of the difference in the products of the specific heat times a characteristic sound velocity on each side, and $\bar{g}$ is a surface conductance, with units 1/$\Omega$-m$^2$.  (We reserve $g$ for the g-factor of the charge carriers; for metals $g\approx-2$.) By the Onsager principle (assumed to apply at surfaces as well as in bulk), $L'_{sn} = L'_{ns}$.  Thus, the total rate of entropy across the surface region is
\begin{equation}
\begin{split}
{\cal{S}}_I &\approx \frac{h_{K}}{T^{2}} \left( \Delta T \right)_I^2 + \frac{\bar{g}}{e^2T} \left( \Delta \tilde{\mu} \right)_I^2 + 2 \frac{L'_{sn}}{T} \left( \Delta \tilde{\mu} \right)_I \left( \Delta T \right)_I.
\label{SurfaceEntropySingle} 
\end{split}
\end{equation}
The condition ${\cal S}_{I}\ge0$ implies that $h_{K} \ge0$, $\bar{g} \ge0$, and $L_{sn}^{'2}\le (\bar{g} h_{K})/(e^2 T)$.  

\subsection{Estimates}
We consider a metal-metal interface, for which $\tilde{\mu}\approx -eV$.  For characteristic values of current density \cite{Katine} ($J \approx 10^{12} $ A/m$^2$) and surface conductance \cite{BassPratt07} ($\bar{g} \approx 10^{15} $ $1/\Omega$-m$^{2}$), a characteristic potential difference across the interface is $(\Delta V)_I \approx 10^{-3}$ V.
Then, when $(\Delta T)_{I}=0$ the appropriate part of $T{\cal S}_{I}$ gives a rate of heating per unit area of
\begin{equation}
{\cal R}_I = T{\cal S}_{I}=\frac{\bar{g}}{e^2} (\Delta \tilde{\mu})_I^2 \approx \bar{g} (\Delta V)_I^2 \approx 10^9 \frac{{\rm W}}{{\rm m}^2}.
\label{JouleI}
\end{equation}

We may also apply this to a multilayer system where interference effects between the layers, and where mean-free paths connecting them, can be neglected \cite{ValetFert}.  At a separation $s$ of 100~nm between layers the net effect of the interfaces corresponds to a bulk conductance of $gs\approx 10^{8}/\Omega$-m and the rate of heating per unit volume (from the values above) is approximately $10^{16}$W/m$^3$.  

As already discussed, there is entropy production because of heat flow, but there is no heating rate associated with heat flux, because the energy is already in the form of heat.  For an interface across which there is only a temperature jump, the rate of entropy production is given by
\begin{equation}
{\cal{S}}_I \approx \frac{h_{K}}{T^{2}} \left( \Delta T \right)_I^2.  
\label{SurfaceEntropyHeat} 
\end{equation}
Nevertheless, there is an {\it apparent} heating rate, whose value we now determine.  Typical values for thermal boundary resistance \cite{SwartzPohl} ($R_{K} = h_{K}^{-1} \approx 2 \times10^{-3}$K-m$^2$/W at $T=1$K for Rh:Fe on Al$_2$O$_3$) and energy flux ($J_\varepsilon \approx 10^{-7} $W/m$^2$) give a value for the temperature difference across the interface of $(\Delta T)_I \approx 2\times10^{-2}$K, so $(\Delta T)_{I}/T\approx 0.02$.  Then
\begin{equation}
T{\cal{S}}_I =\frac{h_{K}}{T} (\Delta T)_I^2 \approx 0.2 \frac{{\rm W}}{{\rm m}^2}.
\end{equation}
This apparent heating rate is about ten orders of magnitude smaller than for the example of a true surface heating rate due to the electrochemical potential gradient, given above in \eqref{JouleI}.  

\subsection{Entropy Production Rates: Bulk {\it vs} Surface}
We now consider the conditions under which the interface entropy production rate ${\cal S}_{I}$ can dominate over the near-surface space-integral ${\cal S}_{B}$ of the bulk entropy production rate ${\cal S}_{s}$.   For simplicity we consider only carrier flow in the $x$-direction, with no cross-terms, so $L_{sn}\approx 0$.  

Three characteristic lengths are associated in this problem: one associated with transport (a mean-free path $l$), a characteristic sample size ($d$), and a distance over which there is an interface adjustment ($a$).  

On equating the bulk and surface carrier currents $\mu j^n$ and using $\partial_{x}\tilde{\mu}\sim \left( \Delta \tilde{\mu} \right)_{B}/d$ we find
\begin{equation}
\left(\Delta \tilde{\mu} \right)_I \approx \frac{\sigma}{hd} \left( \Delta \tilde{\mu} \right)_B.
\end{equation}
With this result, \eqref{R_B2}, and \eqref{SurfaceEntropySingle}, 
for $(\Delta T)_{I}=0$ the integrated bulk rate of entropy production near the surface (per unit area) is on the order of 
  \begin{equation}
  {\cal S}_{B} \sim a \sigma  \frac{\left( \Delta\tilde{\mu} \right)^2_B}{d^{2}}  \sim a\frac{\bar{g}^{2}}{\sigma} ( \Delta \tilde{\mu})^2_I\sim \frac{a\bar{g}}{\sigma}{\cal{S}_{I}}.
  \label{heat-bulk/surf}
  \end{equation}
Thus, when $a\bar{g}/\sigma\gg1$ (good electrical matching), the contributions from the bulk electrochemical potential gradients dominate those near the surface.  On the other hand, when $a\bar{g}/\sigma\ll1$ (poor electrical matching), the contribution from the surface jump in electrochemical potential dominates.

\section{Two-Carrier Systems -- Spin}
\label{Spin}
For itinerant magnets the theory should include two carriers.  A specific case would be the interface between a metal and a magnetic material, where spin-up and spin-down electrons have different electrochemical potentials on each side of the interface.  
For simplicity we consider that the magnetization direction $\hat{M}$ is fixed, and that, for the two adjacent materials, $\hat{M}$ is aligned parallel ($\hat{M}_{1}, \hat{M}_{2} \rightarrow \hat{M}$) or antiparallel ($\hat{M}_{1}\rightarrow\hat{M}$ and $\hat{M}_{2}\rightarrow -\hat{M}$).  (Because we do not consider bulk or surface spin-transfer torque \cite{Saslow2007}, we cannot consider non-collinear $\hat{M}_{1} $ and $\hat{M}_{2}$.) 
Near a magnet, with no external field but with current flow, even a nonmagnetic material can develop a nonequilibrium magnetization (also known as spin accumulation).  With a somewhat different notation, and including terms associated with $\partial_{i}\hat{M}$, for the bulk many of these results (but not surface heating) have been derived previously \cite{Saslow2007}.  Transport of spin across surfaces was considered by a number of authors, but they did not consider heating rates \cite{JohnsonSilsbee,vSonvKempWyder,ValetFert}.

Before introducing the thermodynamics a few definitions are needed.  First, the theory employs the ``magnetization potential'' $-\vec{H}^{*}$.  $\vec{H}^{*}$ itself is given by the difference between the external fields (magnetic, anisotropy, dipole) and the internal field due to exchange.  In equilibrium $\vec{H}^{*}=\vec{0}$.\footnote{There is some ambiguity in the definition of the external fields and the internal fields, but there is no ambiguity in the definition of $\vec{H}^{*}$.  For example, the lattice anisotropy and the dipole fields depends on the magnetization, and for that reason can be considered to be internal or external.  The uniform exchange field is certainly internal, but the non-uniform exchange field might be considered internal or external.  Only the applied magnetic field and the internal exchange field should be uniquely considered external and internal. }  
The  {\it chemical} potentials are denoted by $\mu_{\uparrow,\downarrow}$, and are determined, for example, from energy band theory.  The {\it electrochemical} potentials are denoted by $\tilde{\mu}_{\uparrow,\downarrow}$, and satisfy $\tilde{\mu}_{\uparrow,\downarrow}=\mu_{\uparrow,\downarrow}-eV$.  Finally, the {\it magnetoelectrochemical} potentials are denoted by $\bar{\mu}_{\uparrow,\downarrow}$,\footnote{The term {\it magnetization potential} is employed by \cite{JohnsonSilsbee} to denote $-H^*$.  They note that it would be a term in the {\it magneto-electro-chemical potential,} for which we use  $\bar{\mu}_{\uparrow\downarrow}$.} 
and satisfy 
\begin{equation}
\bar{\mu}_{\uparrow,\downarrow} =  \tilde{\mu}_{\uparrow, \downarrow} \pm (\gamma \hbar/2) \mu_{0} \vec{H}^* \cdot \hat{M}.
\label{magelchempot}
\end{equation}
Here, $\mu_{0}$ is the permeability of free space, and the gyromagnetic ratio $\gamma = |g| \mu_B /\hbar $, where $g$ is the electron spin g-factor and $\mu_B$ is the Bohr magneton.  In equilibrium $\bar{\mu}_{\uparrow}=\bar{\mu}_{\downarrow}$.  Recall that \cite{JohnsonSilsbee} does not consider changes in the chemical potential $\mu_{\uparrow,\downarrow}$.\footnote{Although Johnson and Silsbee \cite{JohnsonSilsbee} formally do not consider changes in the chemical potential, in fact such changes are needed very near the interfaces, in order that the actual voltage (as opposed to the extrapolated voltage) be continuous across the interface.  This is relevant to Fig.~5 of their Appendix.  To make the voltage continuous at the interface requires charges within a screening length of each side of the interface.  Each side has its own screening length, and for a given  side, if the total screening charge (per unit area) is known, then the voltage drop near the surface of that side is known.  The total surface charge per unit area is determined by the difference in the electric fields on each side of the interface.  The distribution of the charges between the two sides is  determined by the requirement that the voltage be continuous. } 

\subsection{Rate of Entropy Production}
With these definitions, the bulk energy density, bulk number densities for spin up ($n_\uparrow$) and spin down ($n_\downarrow$) electrons, and bulk entropy density are related by 
\begin{equation}
d\varepsilon=Tds + \bar{\mu}_\uparrow dn_\uparrow + \bar{\mu}_\downarrow dn_\downarrow.
\label{2depsilon}
\end{equation}
Here we neglect a term $\sim \vec{H} \cdot d\hat{M}$ because we restrict ourselves to the longitudinal response.  

The relations for energy density \eqref{e_eq} and entropy density \eqref{R/T} still apply.  
Relations for the (non-conserved) number flux densities are given by
\begin{eqnarray}
\frac{\partial n_\uparrow}{\partial t} +\partial_{i} j_{\uparrow i} &=& S_{\uparrow} \rm{,}
\label{nup_eq} \\
\quad \frac{\partial n_\downarrow}{\partial t} +\partial_{i} j_{\downarrow i} &=& S_{\downarrow}=- S_{\uparrow} \rm{,}
\label{ndown_eq}
\end{eqnarray}
where $S_{\uparrow}$ is the rate at which spin-down electrons flip to spin-up electrons.  These forms ensure that the total number current, 
\begin{equation}
J^n_{i} \equiv (j_{\uparrow {i}} + j_{\downarrow {i}}),
\label{Jn}
\end{equation}
is conserved, since
\begin{equation}
\frac{\partial \left( n_\uparrow + n_\downarrow \right)}{\partial t} + \partial_{i} J^n_{i} = 0.
\label{charge_eq}
\end{equation}
However, the dimensionless spin current,
\begin{equation}
J^\sigma_i \equiv j_{\uparrow i} - j_{\downarrow i}\rm{,}
\label{Jsigma}
\end{equation} 
is not conserved:
\begin{equation}
\frac{\partial \left( n_\uparrow - n_\downarrow \right)}{\partial t} +\partial_{i} J^\sigma_{i} = 2 S_{\uparrow}.
\label{sigma_eq}
\end{equation}

By equations \eqref{e_eq}, \eqref{R/T}, \eqref{nup_eq} and \eqref{ndown_eq}, we have
 \begin{equation}
 \begin{split}
 0 \leq T{\cal S}_{s} =& \partial_{i} \left( -j^\varepsilon_{i} + T j^s_{i} + \bar{\mu}_\uparrow j_{\uparrow i} + \bar{\mu}_\downarrow j_{\downarrow i}  \right)\\
  & - j^s_{i} \partial_{i} T - j_{\uparrow i} \partial_{i}\bar{\mu}_\uparrow - j_{\downarrow i} \partial_{i}\bar{\mu}_\downarrow - \left(\bar{\mu}_\uparrow - \bar{\mu}_\downarrow \right) S_{\uparrow}\rm{.}
 \end{split}
 \end{equation}
 Again, in order to ensure that the entropy does not decrease, the divergence must be zero, so we take
 \begin{equation}
 j^\varepsilon_{i} = T j^s_{i} + \bar{\mu}_\uparrow j_{\uparrow i} + \bar{\mu}_\downarrow j_{\downarrow i}\rm{.}
 \label{2jRelation}
 \end{equation}
 Then,
 \begin{equation}
T{\cal S}_{s} =  - j^s \partial_i T - j_\uparrow \partial_i \bar{\mu}_\uparrow - j_\downarrow \partial_i \bar{\mu}_\downarrow - \left(\bar{\mu}_\uparrow - \bar{\mu}_\downarrow \right) S_{\uparrow} .
 \label{2R1}
 \end{equation}
Not only does each partial current and the heat current contribute to the rate of heating, but there is also a contribution from spin-flip, involving the difference in magnetoelectrochemical potential for up and down spins.  For conserved spin current (i.e., $\alpha=0$), eq.~\eqref{2R1} agrees with eq.~(53) of \cite{JohnsonSilsbee}.  See Table~\ref{JSTable} for a term-by-term comparison.

\subsection{Bulk Fluxes and Rate of Entropy Production}
In the bulk, from equation \eqref{2R1}, we take the linearized flux densities to be 
\begin{align}
j^s_{i} =& - \frac{\kappa}{T} \partial_{i} T - L_{s \uparrow} \partial_{i} \bar{\mu}_\uparrow - L_{s \downarrow} \partial_{i} \bar{\mu}_\downarrow ,\label{jsSPIN}\\
j_{\uparrow i}=& - L_{\uparrow s} \partial_{i} T - \frac{\sigma_\uparrow}{e^2} \partial_{i} \bar{\mu}_{\uparrow} - L_{\uparrow \downarrow} \partial_{i} \bar{\mu}_{\downarrow} ,\label{jup}\\
j_{\downarrow i}=& - L_{\downarrow s} \partial_{i} T - L_{\downarrow \uparrow} \partial_{i} \bar{\mu}_{\uparrow} - \frac{\sigma_\downarrow}{e^2} \partial_{i}\bar{\mu}_{\downarrow} ,\label{jdown}
\end{align}
where $\sigma_\uparrow$ and $\sigma_\downarrow$ are electrochemical conductivities of up spins and down spins, respectively.  By the Onsager principle, $L_{\uparrow \downarrow} = L_{\downarrow \uparrow}$, $L_{\uparrow s} = L_{s \uparrow} $, and $ L_{s \downarrow} = L_{\downarrow s} $.  Further, to ensure the non-negativity of \eqref{2R1} even in the absence of gradients of intensive variables, $S_{\uparrow}$ is driven by the difference in electrochemical potentials
\begin{equation}
S_{\uparrow}=-\alpha \left(\bar{\mu}_{\uparrow}-\bar{\mu}_{\downarrow} \right), 
\label{S}
\end{equation}
where $\alpha$ is proportional to a characteristic spin-flip rate.

We now expand the magnetoelectrochemical potentials as
\begin{align}
\bar{\mu}_{\uparrow} = {\bar{\mu}_{\uparrow}}^{(0)} + \frac{\partial \bar{\mu}_{\uparrow}}{\partial n_{\uparrow}} \delta n_{\uparrow}, \quad \bar{\mu}_{\downarrow} = {\bar{\mu}_{\downarrow}}^{(0)} + \frac{\partial \bar{\mu}_{\downarrow}}{\partial n_{\downarrow}} \delta n_{\downarrow}. 
\end{align}
Using ${\bar{\mu}_{\uparrow}}^{(0)} = {\bar{\mu}_{\downarrow}}^{(0)}$ we then have
\begin{align}
S_{\uparrow} = - \alpha \left( C_{\uparrow}\delta n_{\uparrow} - C_{\downarrow}\delta n_{\downarrow} \right),
\end{align}
where $C_{\uparrow} \equiv(\partial \bar{\mu}_{\uparrow}/\partial n_{\uparrow})$ and $C_{\downarrow} \equiv(\partial \bar{\mu}_{\downarrow}/\partial n_{\downarrow})$ are the inverse densities of states.  (A similar relation holds in semiconductors, where electrons and holes recombine \cite{KrcmarSas}.)  

We now relate this to $T_{1}$ relaxation of the magnetization $M$ of a uniform system, where 
\begin{equation}
M=-\frac{|g|\mu_{B}}{2}(n_{\uparrow}-n_{\downarrow}). 
\label{M}
\end{equation}
Recall that $g$ is the charge carrier g-factor. 
Particle conservation gives $\delta n_{\downarrow}=-\delta n_{\uparrow}$, so $\delta M=-|g|\mu_{B}\delta n_{\uparrow}$ and $S_{\uparrow} = - \alpha ( C_{\uparrow}+C_{\downarrow})\delta n_{\uparrow} $.   Then \eqref{sigma_eq}, with $\partial_{i}J^{\sigma}_{i}$ neglected, yields 
\begin{equation}
\frac{\partial M}{\partial t}=-|g|\mu_{B}\frac{\partial}{\partial t} (\delta {n}_{\uparrow})=-\frac{M}{T_{1}}, 
\label{M2}
\end{equation}
where $T_{1}^{-1}=\alpha(C_{\uparrow}+C_{\downarrow})$.  

The dissipation function $R=T{\cal S}_{s}$ is given for the bulk by
\begin{equation}
\begin{split}
T{\cal S}_{s} =&   \frac{\kappa}{T} \left( \partial_{i} T \right)^2 + \frac{\sigma_\uparrow}{e^2} \left(\partial_{i} \bar{\mu}_{\uparrow} \right)^2 + \frac{\sigma_\downarrow}{e^2} \left(\partial_{i} \bar{\mu}_{\downarrow} \right)^2 \\
&+ 2 L_{s \uparrow} \left( \partial_{i} \bar{\mu}_{\uparrow}  \right) \left( \partial_{i} T \right) + 2 L_{s \downarrow} \left( \partial_{i} \bar{\mu}_{\downarrow} \right) \left( \partial_{i} T \right)\\
&+ 2 L_{\uparrow \downarrow} \left( \partial_{i} \bar{\mu}_{\uparrow}  \right) \left( \partial_{i} \bar{\mu}_{\downarrow} \right) + \alpha \left(\bar{\mu}_{\uparrow} - \bar{\mu}_{\downarrow} \right)^{2} \rm{.}\label{RB}
\end{split}
\end{equation}
This is sufficiently complex that each term deserves comment.  

The term in $( \partial_{i} T )^2$ is from heat current, the terms in $( \partial_{i} \bar{\mu}_{\uparrow} )^2$ and $( \partial_{i} \bar{\mu}_{\downarrow} )^2$ are from Joule losses of the individual carriers \cite{SaslowCuZn}, the next three are cross-terms, and the last term gives the dissipation due to spin-flip processes.  ${\cal S}_{s}\ge0$ forces various conditions on both the diagonal and the cross-terms, of which the latter usually are small.  
The diagonal terms satisfy $\kappa \geq 0$, $\sigma_{\uparrow} \geq 0$, $\sigma_{\downarrow} \geq 0$, and $\alpha \geq 0$.
Note that it is not the current or spin current (both of them thermodynamic fluxes) that determines the rate of entropy production and heating, but rather the gradients of the magnetoelectrochemical potentials (both of them thermodynamic forces).

Ref.~\cite{JohnsonSpinCal} considers bulk entropy production due to spin accumulation in a non-magnetic material that is adjacent to a ferromagnet. The present work considers the bulk entropy production in both the non-magnetic material and the ferromagnet, as well as the total rate of entropy production at and near the interface (see below).

\subsection{Dissipationless Spin Currents}
In recent years there has been considerable interest in the possibility of spin currents that do not cause dissipation.  Although in \cite{Murakami03,Sinova04} an electric field is applied, Rashba notes that it is possible to have (equilibrium) spin currents, even without an applied electric field \cite{Rashba}.  This situation can be incorporated within the present theory by: 
\begin{itemize}
\item[(a)] letting $j_{\uparrow i}\rightarrow j^{(eq)}_{\uparrow i}+\delta j_{\uparrow i}$ and $j_{\downarrow i}\rightarrow j^{(eq)}_{\downarrow i}+\delta j_{\downarrow i}$ in \eqref{nup_eq} and \eqref{ndown_eq} and the remainder of that subsection, with 
$\partial_{i}j^{(eq)}_{\uparrow i}=\partial_{i}j^{(eq)}_{\downarrow i}=0$; and 

\item[(b)] taking
\eqref{jup} to apply to $\delta j_{\uparrow i}$ and \eqref{jdown} to apply to $\delta j_{\downarrow i}$. 
\end{itemize}
Because the rate of entropy production involves the thermodynamic forces, \eqref{RB} still applies.  For the rate of heating, one should apply \eqref{2R1} with the $ j_i^s \partial_i T $ term omitted.

\subsection{Surface Rate of Entropy Production}
Consider flow along $x$, so we may drop the directional indices on the fluxes.  Then, rewriting \eqref{2R1} with \eqref{Jn} and \eqref{Jsigma} gives
\begin{align}
T{\cal S}_{s} =&  - j^s \partial_x T - \frac{1}{2}J^n \partial_x \left( \bar{\mu}_{\uparrow} + \bar{\mu}_{\downarrow} \right) \notag\\ 
&- \frac{1}{2}J^\sigma \partial_x \left( \bar{\mu}_{\uparrow} - \bar{\mu}_{\downarrow} \right) - \left(\bar{\mu}_{\uparrow} - \bar{\mu}_{\downarrow} \right) S_{\uparrow}.
\label{2R2}
\end{align}
In steady-state $J^n$ is constant across the interface region, but $J^\sigma$ is not.  Moreover, $j^s$ is not obviously a near-conserved quantity, unlike for a single carrier.  However, since $j^\varepsilon$ is conserved, we use \eqref{2jRelation} to write
\begin{equation}
j^s = \frac{1}{T} j^\varepsilon - \frac{1}{2} J^n \left( \frac{\bar{\mu}_{\uparrow}}{T} + \frac{\bar{\mu}_{\downarrow}}{T} \right) - \frac{1}{2} J^\sigma \left( \frac{\bar{\mu}_{\uparrow}}{T}- \frac{\bar{\mu}_{\downarrow}}{T} \right).
\end{equation}
With both $J^{\sigma}$ and the difference in magnetoelectrochemical potentials considered to be first order in 
small deviations from equilibrium, the last term is second order.  Thus $j^s$ actually is a  near-conserved quantity.

Integrating the volume rate of heating \eqref{2R2} over the interface region yields
\begin{align}
T{\cal S}_{I} =&  -j^s \left(\Delta T \right)_I  - \frac{1}{2} J^n \left( \left( \Delta \bar{\mu}_{\uparrow} \right)_I+ \left(\Delta \bar{\mu}_{\downarrow} \right)_I \right) \notag\\
& - \frac{1}{2} \int_I dx  J^\sigma \partial_x \left( \bar{\mu}_{\uparrow} - \bar{\mu}_{\downarrow} \right)- \int_I dx  \left( \bar{\mu}_{\uparrow} - \bar{\mu}_{\downarrow} \right) S_{\uparrow} .
\label{2R3}
\end{align}
Integrating the third term by parts and using equation \eqref{sigma_eq} with the time-derivative set to zero (steady-state, so that $\partial_x J^\sigma = 2 S$) gives
\begin{equation}
\begin{split}
- &\frac{1}{2} \int_I dx  J^\sigma \partial_x \left( \bar{\mu}_{\uparrow} - \bar{\mu}_{\downarrow} \right) \\
&= \int_I dx \left( \bar{\mu}_{\uparrow} - \bar{\mu}_{\downarrow} \right) S_{\uparrow}
- \frac{1}{2} \Delta \left( J^\sigma \left( \bar{\mu}_{\uparrow} - \bar{\mu}_{\downarrow} \right) \right)_I   .
\label{2R4}
\end{split}
\end{equation}
The first term on the right-hand-side (RHS) of \eqref{2R4} cancels the last term on the RHS of \eqref{2R3}, and the second term is evaluated on each side of the interface.  
Then \eqref{2R3} becomes 
\begin{equation}
\begin{split}
T{\cal S}_{I} =& -j^s  \left(\Delta T \right)_I - \frac{1}{2} J^n \left( \left( \Delta \bar{\mu}_{\uparrow} \right)_I+ \left(\Delta \bar{\mu}_{\downarrow} \right)_I \right)\\
& - \frac{1}{2} \Delta \left( J^\sigma \left( \bar{\mu}_{\uparrow} - \bar{\mu}_{\downarrow} \right) \right)_I \rm{.}
\label{2R_S1}
\end{split}
\end{equation}
The last term appears to be well-defined, but because $J^{\sigma}$ is not conserved, it is not clear how to interpret this unambiguously. 
Nevertheless, if the spin diffusion length, over which up and down spins flip, is sufficiently long relative to the surface region, so that one can measure $J^{\sigma}$ within a spin diffusion length of the interface, then this last term should be clearly defined (otherwise it may not be measurable).\footnote{Because $\bar{\mu}_{\uparrow,\downarrow}$ and $J^{\sigma}$  vary exponentially within each material's spin-diffusion length from the interface, there are four intervals over which $T {\cal S}_{I}$ of \eqref{2R_S1} is clearly defined.   Taking the edges of the intervals to be either much closer or much farther than the spin-diffusion lengths from the interface yields four possible intervals.}  
In what follows we will assume that this holds.  For conserved spin current, eq.~\eqref{2R_S1} agrees with eq.~(44) of \cite{JohnsonSilsbee}.  See Table~\ref{JSTable} for a term-by-term comparison. 

\subsection{Surface Fluxes and Rate of Entropy Production}
In the spirit of irreversible thermodynamics, the entropy and the spin up and spin down number fluxes can be linearized in differences in the appropriate intensive thermodynamic quantities across the interface, so 
\begin{eqnarray}
j^s &=& - \frac{h_{K}}{T} \left( \Delta T \right)_I - L'_{s \uparrow} \left(\Delta \bar{\mu}_{\uparrow}\right)_I - L'_{s \downarrow} \left(\Delta \bar{\mu}_{\downarrow}\right)_I ,\label{js2}\\
j_{\uparrow }&=& - L'_{\uparrow s} \left(\Delta T\right)_I - \frac{g_\uparrow}{e^2} \left(\Delta \bar{\mu}_{\uparrow}\right)_I - L'_{\uparrow \downarrow} \left(\Delta \bar{\mu}_{\downarrow}\right)_I ,\\
j_{\downarrow }&=& - L'_{\downarrow s} \left(\Delta T\right)_I- L'_{\downarrow \uparrow} \left(\Delta \bar{\mu}_{\uparrow}\right)_I - \frac{g_\downarrow}{e^2} \left(\Delta \bar{\mu}_{\downarrow} \right)_I. \label{jdownISPIN}
\end{eqnarray}
Here $g_\uparrow$ and $g_\downarrow$ are surface conductances of spin up and spin down particles, and by the Onsager principle $L'_{\uparrow \downarrow} = L'_{\downarrow \uparrow}$, $L'_{\uparrow s} = L'_{s \uparrow} $, and $ L'_{s \downarrow} = L'_{\downarrow s} $.  For a calculation of a spin-dependent interfacial surface resistance, see Ref.~\cite{StilesPenn}.  

From \eqref{Jn} and \eqref{Jsigma} the total number current and spin current can be written as
\begin{align}
J^n = - \left( L'_{\uparrow s} + L'_{\downarrow s} \right) \left(\Delta T\right)_I - \left( \frac{g_\uparrow}{e^2} + L'_{\downarrow \uparrow} \right) \left( \Delta \bar{\mu}_{\uparrow} \right)_I \notag\\
- \left(  \frac{g_\downarrow}{e^2}  + L'_{\uparrow \downarrow}\right) \left(\Delta \bar{\mu}_{\downarrow}\right)_I ,
\label{Jn2}
\end{align}
\begin{align}
J^\sigma = - \left( L'_{\uparrow s} - L'_{\downarrow s} \right)  \left(\Delta T\right)_I- \left(\frac{g_\uparrow}{e^2} - L'_{\downarrow \uparrow} \right) \left(\Delta \bar{\mu}_{\uparrow}\right)_I \notag\\
 + \left( \frac{g_\downarrow}{e^2}- L'_{\uparrow \downarrow} \right) \left(\Delta \bar{\mu}_{\downarrow} \right)_I.
 \label{Jsig1}
\end{align}
Substitution of the currents in \eqref{js2} and \eqref{Jn2} into \eqref{2R_S1} yields
\begin{align}
T{\cal S}_{I}& =- \frac{1}{2} \Delta \left( J^\sigma \left( \bar{\mu}_{\uparrow} - \bar{\mu}_{\downarrow} \right) \right)_I+\frac{h_{K}}{T}  \left( \Delta T \right)_I^2 \notag\\
&+ \frac{1}{2}\left( \frac{g_\uparrow}{e^2} + L'_{\downarrow \uparrow}\right)\left(\Delta \bar{\mu}_{\uparrow} \right)_I^2 + \frac{1}{2}\left( \frac{g_\downarrow}{e^2} + L'_{\downarrow \uparrow}\right)\left(\Delta \bar{\mu}_{\downarrow} \right)_I^2\qquad \notag\\ 
&+ \frac{1}{2} \left( 3 L'_{s \uparrow} + L'_{\downarrow s} \right) \left(\Delta \bar{\mu}_{\uparrow} \right)_I\left(\Delta T \right)_I \notag\\
&+ \frac{1}{2} \left( 3 L'_{s \downarrow} + L'_{\uparrow s} \right) \left(\Delta \bar{\mu}_{\downarrow} \right)_I \left(\Delta T \right)_I \notag\\
&+ \frac{1}{2} \left( \frac{g_\uparrow + g_\downarrow}{e^2} + 2 L'_{\downarrow \uparrow} \right) \left(\Delta \bar{\mu}_{\uparrow} \right)_I \left(\Delta \bar{\mu}_{\downarrow} \right)_I.
\label{2R_S2}
\end{align}

If the spin current $J^\sigma$ is approximately uniform near the surface, then use of \eqref{Jsig1} gives that \eqref{2R_S2} simplifies to
\begin{align}
T{\cal S}_{I} =& \frac{h_{K}}{T}  \left( \Delta T \right)_I^2 + \frac{g_\uparrow}{e^2}\left(\Delta \bar{\mu}_{\uparrow} \right)_I^2 + \frac{g_\downarrow}{e^2}\left(\Delta \bar{\mu}_{\downarrow} \right)_I^2\notag\\ 
&+ 2 L'_{s \uparrow}  \left(\Delta \bar{\mu}_{\uparrow} \right)_I \left(\Delta T \right)_I + 2 L'_{s \downarrow}  \left(\Delta \bar{\mu}_{\downarrow} \right)_I \left(\Delta T \right)_I \qquad \notag\\
&+ 2 L'_{\downarrow \uparrow} \left(\Delta \bar{\mu}_{\uparrow} \right)_I \left(\Delta \bar{\mu}_{\downarrow} \right)_I .
\label{2R_SFinal}
\end{align}
as in \cite{JohnsonSilsbee}, which gives an approximation for each of the coefficients.

\subsection{Comparison of Electric and Spin Current Heating}

Eq.~\eqref{2R_SFinal} permits a comparison of surface heating due to electric current (equivalently, due to a voltage jump across the surface region) with heating due to spin current (equivalently, due to the difference in $\mu_{0} \vec{H}^*\cdot \hat{M}$ across the surface region).  We consider a metal-metal interface where temperature is uniform, spin is conserved across the interface, and $g_\uparrow \approx g_\downarrow \approx \bar{g}/2 $.  

For the purposes of estimation, we neglect the chemical potentials $\mu_\uparrow$ and $\mu_\downarrow$.  The limitations of this approximation are discussed above.  Then, \eqref{magelchempot} gives
\begin{align}
\left(\Delta \mu_{\uparrow, \downarrow}\right)_I^2  \approx& -e^2 (\Delta V)_I^2 + \frac{\gamma^2 \hbar^2}{4} \mu_{0}^2 (\Delta H^*)_I^2 \notag\\
&\mp \gamma \hbar e \mu_0 (\Delta V)_I (\Delta H^*)_I.
\label{muestim}
\end{align}
Here we define $H^* \equiv \vec{H}^* \cdot \hat{M}$.  
To find the heating due to $(\Delta V)_I$ and $\mu_{0} (\Delta H^*)_I$, we substitute \eqref{muestim} into the second and third terms of \eqref{2R_SFinal}.

Suppose (as above) that the surface region is characterized by surface conductivity $\bar{g} = 10^{15}$~1/$\Omega$-m$^2$, and voltage difference $(\Delta V)_I = 10^{-3}$~V.  Neglecting cross-terms, the rate of heating, per unit area, due {\it only} to the voltage difference is given by
\begin{align}
{\cal R}_{I}^{elec}=T {\cal S}_{I}^{elec} \approx  \bar{g} \left[ (\Delta V)_I \right]^2 \approx 10^9 ~\frac{{\rm W}}{{\rm m}^2}.
\end{align}
(Cancellation of the term proportional to $(\Delta V)_I (\Delta H^*)_I $ is only approximate.)

On the other hand, neglecting cross-terms, the rate of heating due {\it only} to spin current is given by
\begin{align}
{\cal R}_{I}^{spin}=T {\cal S}_{I}^{spin} \approx& \frac{\bar{g}}{e^2} \left[\frac{\gamma \hbar }{2} \mu_{0} (\Delta H^*)_I \right]^2  \notag\\
=& \frac{\bar{g}}{4 e^2} \left[|g| \mu_B \mu_{0} (\Delta H^*)_I \right]^2.
\end{align}
For $|g| \approx 2$ and $\mu_B \approx 5.8 * 10^{-5}$~eV/T, 
\begin{align}
{\cal R}_{I}^{spin}\approx 3.4 * 10^{6}  \left[ \mu_{0} (\Delta H^*) \right]^2\frac{{\rm W}}{{\rm T}^{2} {\rm -m}^2}.
\end{align}
Thus, a $\mu_{0} (\Delta H^*)_I \approx 20 $~T gives about the same heating as a voltage difference of $(\Delta V)_I = 10^{-3}$~V.

\section{Summary and Conclusions}
\label{Conclusion}
We have applied the methods of irreversible thermodynamics to study the rate of entropy and heat production in bulk and associated with an interface, first for an ordinary conductor (this includes insulators) and then for a conducting magnet whose magnetization direction $\hat{M}$ is fixed.  In addition to magnetic metals, the present results apply to magnetic insulators and semiconductors due to the incorporation of the chemical potential, neglected previously.  We also show how equilibrium spin currents can be included in the present theory. This is relevant to heating that takes place in nanoelectronic systems and spintronic systems, especially in multilayers.  

We conclude by noting that the present work is related to recent work on ``spin caloritronics,'' whereby heat currents can cause spin currents and spin currents can cause heat currents \cite{Hatami,SpinCal, YuAnsermet10}.  The former, known as the spin-Seebeck effect, has recently been measured\cite{UchidaPy,UchidaInsul,AwschMyers} using the inverse spin Hall effect, and in one case\cite{AwschMyers} displays a profile that is associated with spatially exponential decay away from the heat input and output leads \cite{Bauer1,SearsSasSSE}.

\section*{Acknowledgements}
We would like to acknowledge conversations with C. R. Hu, V. E. Pokrovsky, and the support of the DOE through grant DE-FG02-06ER46278.

\appendix*

\section{Entropy-Maximum-Based Variables}
\label{Entropy-BasedAppendix}

One may write the increase in entropy in terms of entropy-maximum-based intensive variables (e.g., $1/T$, $-\mu/T$).  This is most easily shown by writing the time derivative of the entropy density as
\begin{align}
\dot{s} = \sum_{k} \frac{\partial s}{\partial x_k} \dot{x}_k,
\label{dotS-a}
\end{align}
where the index $k$ here is used to denote densities $x_k$ of the entropy-maximum-based extensive variables.  These include the energy density $\epsilon$, which is conjugate to $1/T$.  We take an equation of motion that includes both flux $J_i^k$ and source ${\cal S}_k$ (${\cal S}_k=0$ if $x_k$ is conserved, but for spin-flip scattering it is important): 
\begin{align}
\dot{x}_k = -\partial_i J_i^k + {\cal S}_k,
\end{align}
Then \eqref{dotS-a} gives
\begin{align}
\dot{s}=-\partial_i J_i^s + {\cal S}_s = -\sum_{k} \frac{\partial s}{\partial x_k} \partial_i{J_i^k} + \sum_{k} \frac{\partial s}{\partial x_k}{\cal S}_k,
\label{dotS2-a}
\end{align}
Solving for ${\cal S}_{s}$ (which must be non-negative) and rewriting so that there is an independent gradient term yields
\begin{align}
0 \le {\cal S}_s =& \partial_i \left[ J_i^s - \sum_{k} \left(\frac{\partial s}{\partial x_k} J_i^k\right) \right] \notag\\
&+ \sum_{k} {J_i^k} \partial_i \left( \frac{\partial s}{\partial x_k}\right) + \sum_{k} {\cal S}_k\frac{\partial s}{\partial x_k}.
\end{align}
Since ${\cal S}_S \ge 0$ but the divergence may take either sign, the argument of the divergence must be zero (modulo a curl, which we neglect).  This relates the entropy flux to the other fluxes, but does not determine them.  

To obtain the other fluxes, we set the divergence to zero: 
\begin{align}
0 \le {\cal S}_s = \sum_{k} {J_i^k} \partial_i \left( \frac{\partial s}{\partial x_k}\right) + \sum_{k} {\cal S}_k\frac{\partial s}{\partial x_k}.
\end{align}
This is the entropy-maximum variable analog of \eqref{R2} and \eqref{2R1}, which are in energy-minimum variables.  Only the first term would appear if the extensive quantities were conserved; the second term represents the effect of sources.  To obtain the irreversible thermodynamics one can expand the unknown sources and fluxes in terms of the presumably known $\partial s/\partial x_k$'s and their gradients. 

\section{Sound Waves}
\label{SoundAppendix}
Consider a sound wave.  Let subscript $0$ denote an equilibrium value.  
With $\sigma = s/\rho$, the time-averaged change in entropy density due to a sound wave is
\begin{align}
\overline{s-s_0} =&\frac{\partial s}{\partial \sigma} \overline{\delta \sigma} + \frac{\partial s}{\partial \rho} \overline{\delta \rho}+ \frac{1}{2}\frac{\partial^2 s}{\partial \sigma \partial \rho} \overline{(\delta \sigma)(\delta \rho)} + \frac{1}{2}\frac{\partial^2 s}{\partial \sigma^2} \overline{(\delta \sigma)^2}\notag\\
& + \frac{1}{2}\frac{\partial^2 s}{\partial \rho^2} \overline{(\delta \rho)^2} + \frac{1}{2}\frac{\partial s}{\partial (v^2)} \overline{ v^2} + \dots.
\label{SoundWaves}
\end{align}
For sound in the long-wavelength limit $\sigma$ is constant, so $d\sigma=0$.  Moreover, $\overline{\delta \rho}=0$.  Therefore the first four terms are zero.  Further, the entropy is determined from the rest frame, and thus has no velocity-dependence, so the sixth term is also zero.  Finally, since $s = \rho \sigma$ and each derivative is taken with the other variables held constant, the fifth term also is zero:
\begin{align}
\left. \frac{\partial^2 s}{\partial \rho^2} \right|_\sigma =& \left. \frac{\partial \sigma}{\partial \rho} \right|_\sigma=0.
\end{align}
Therefore, to second order in fluctuations, a sound wave has zero entropy; its energy is solely in the form of potential energy and kinetic energy.   

\section{Notational Variation}
\label{NotationAppendix}
Callen (see eq.~(17.3) of Ref.~\cite{Callen}) uses gradients of entropic variables (e.g., $1/T$ and $-\mu/T$) for generalized thermodynamic forces.  
Except for Appendix \ref{Entropy-BasedAppendix}, the present work uses gradients of energy-minimum variables (e.g., $T$, $\mu$).  Morse (see eqs.~(8.36) and (8.48) of Ref.~\cite{Morse}) shows the results of both approaches.  Johnson and Silsbee (see eq.~(12) of Ref.~\cite{JohnsonSilsbee}) use neither convention, and have gradients of $1/T$ and voltage $V$ for generalized forces.  

These differences make comparison complex.  Table~\ref{JSTable} compares the quantities of the present work to those of Ref.~\cite{JohnsonSilsbee}.  These authors take the electron charge as $e=-|e|$, $\beta=\mu_B$ as the Bohr magneton, and the spin g-factor as positive. We take the gyromagnetic ratio to be$\gamma = |g| \mu_B /\hbar $.  
Recall that Ref.~\cite{JohnsonSilsbee} neglects the chemical potentials ($\mu$, $\mu_{\uparrow}$ and $\mu_{\downarrow}$) relative to $-eV$. Thus Table~\ref{JSTable} applies for $\tilde{\mu} \approx -eV$. 

\begin{table*}[h!]
\caption{The relationship between quantities in the present work and in Ref.~\cite{JohnsonSilsbee}.  We take $e>0$ in this table and in the present work, whereas Ref.~\cite{JohnsonSilsbee} takes $e<0$. }
\begin{center}
\begin{tabular}{l | c c|c c}
\hline \hline
& \multicolumn{2}{c|}{Bulk (Continuous System) Quantities} & \multicolumn{2}{c}{Surface (Discrete System) Quantities}\\ \cline{2-5}  
&~~Present  ~~ & ~~ Ref.~\cite{JohnsonSilsbee}~~~ & ~~Present~~ & ~~ Ref.~\cite{JohnsonSilsbee}~~~ \\ \hline
Entropy Production & ${\cal S}_s$ & $\dot{S}_{univ}$ & ${\cal S}_I$ & $\dot{S}_{univ}$\\ [1ex]
Entropy Flux & $j^s$ & $\displaystyle \frac{J_Q}{T}$ & $j^s$ & $\displaystyle \frac{I_Q}{T}$\\ [2ex] 
Charge-Carrier Flux & $j^n$, $J^n$ &$\displaystyle - \frac{J_q}{e} $  & $j^n$, $J^n$ &$\displaystyle - \frac{I_q}{e} $\\ [2ex] 
Spin Flux & $J^\sigma$ &$\displaystyle   -\frac{2 J_M}{\gamma \hbar} $ & $J^\sigma$ &$\displaystyle   -\frac{2 I_M}{\gamma \hbar} $\\ [2ex] 
Thermal Conductivity & \multirow{2}{*}{$\kappa$ }& \multirow{2}{*}{$\displaystyle \frac{L_{22}}{T^2}$ }& \multirow{2}{*}{$h_{K}$ }& \multirow{2}{*}{$\displaystyle \frac{L_{22}'}{T^2}$}\\
and Conductance &&&& \\ [2ex] 
Electrical Conductivity & \multirow{2}{*}{$\sigma$ }& \multirow{2}{*}{$\displaystyle   \frac{L_{11}}{T} $ }& \multirow{2}{*}{$\bar{g}$ }& \multirow{2}{*}{$\displaystyle   \frac{L_{11}'}{T} $}\\ 
and Conductance &&&& \\[2ex] 
Electrical Conductivity and & \multirow{2}{*}{$\sigma_{\uparrow} $ }&\multirow{2}{*}{$\displaystyle \frac{L_{11}}{4 T} + \frac{e^2 L_{33}}{\gamma^2 \hbar^2 T} + \frac{e L_{31}}{\gamma \hbar T}$ }& \multirow{2}{*}{$g_{\uparrow} $ }& \multirow{2}{*}{$\displaystyle \frac{L_{11}'}{4 T} + \frac{e^2 L_{33}'}{\gamma^2 \hbar^2 T} + \frac{e L_{31}'}{\gamma \hbar T}$}\\ 
 Conductance for $\uparrow$-Spins &&&&\\[2ex] 
Electrical Conductivity and & \multirow{2}{*}{$\sigma_{\downarrow} $  }&\multirow{2}{*}{$\displaystyle \frac{L_{11}}{4 T} + \frac{e^2 L_{33}}{\gamma^2 \hbar^2 T} - \frac{e L_{31}}{\gamma \hbar T}$ }& \multirow{2}{*}{$g_{\downarrow} $ }&\multirow{2}{*}{$\displaystyle \frac{L_{11}'}{4 T} + \frac{e^2 L_{33}'}{\gamma^2 \hbar^2 T} - \frac{e L_{31}'}{\gamma \hbar T}$}\\
 Conductance for $\downarrow$-Spins &&&&\\[2ex]
$s \leftrightarrow n$ Onsager Coefficient & $L_{sn}$ &$\displaystyle   \frac{L_{21}}{e T^2}  $ & $L_{sn}'$ &$\displaystyle   \frac{L_{21}'}{e T^2}  $ \\ [2ex] 
$s \leftrightarrow \,\, \uparrow$ Onsager Coefficient & $L_{s \uparrow} $ &$\displaystyle \frac{L_{32}}{\gamma \hbar T^2} + \frac{L_{12}}{2 e T^2}$ & $L_{s \uparrow}' $ &$\displaystyle \frac{L_{32}'}{\gamma \hbar T^2} + \frac{L_{12}'}{2 e T^2}$\\ [2ex] 
$s \leftrightarrow \,\, \downarrow$ Onsager Coefficient & $L_{s \downarrow} $ &$\displaystyle -\frac{L_{32}}{\gamma \hbar T^2} + \frac{L_{12}}{2 e T^2} $ & $L_{s \downarrow}' $ &$\displaystyle -\frac{L_{32}'}{\gamma \hbar T^2} + \frac{L_{12}'}{2 e T^2} $\\ [2ex] 
$\uparrow \,\, \leftrightarrow \,\, \downarrow$ Onsager Coefficient & $L_{\uparrow \downarrow} $ &$\displaystyle \frac{L_{11}}{4 e^2 T} - \frac{L_{33}}{\gamma^2 \hbar^2 T} $ & $L_{\uparrow \downarrow}' $ &$\displaystyle \frac{L_{11}'}{4 e^2 T} - \frac{L_{33}'}{\gamma^2 \hbar^2 T} $\\ 
 [2ex] \hline \hline
\end{tabular}
\end{center}
\label{JSTable}

\end{table*}

\end{document}